\documentclass{PoS}
\usepackage{xspace}
\usepackage{amstext}
\usepackage{lineno}
\newcommand{\pT}{$p_{\mbox{\tiny T}}$\xspace}
\newcommand{\sNN}{$\sqrt{s_{\mbox{\tiny NN}}}$\xspace}
\newcommand{\s}{$\sqrt{s}$\xspace}

\newcommand{\sA}{\s~=~7~TeV\xspace}
\newcommand{\sU}{\sNN~=~8.16~TeV\xspace}
\newcommand{\sD}{\s~=~13~TeV\xspace}
\newcommand{\sG}{\s~=~5.02~TeV\xspace}
\newcommand{\sC}{\sNN~=~5.02~TeV\xspace}

\title{J/$\psi$ production measurements at mid-rapidity using the ALICE detector at the LHC}

\ShortTitle{J/$\psi$ measurements in ALICE}

\author{\speaker{Ingrid McKibben Lofnes}, on behalf of the ALICE Collaboration\\ 
        Department of Physics and Technology, University of Bergen (Norway)\\
        E-mail: \email{ingrid.mckibben.lofnes@cern.ch}}

\abstract{ Charmonium production is highly sensitive to the hot and dense medium created in (ultra)-relativistic heavy-ion collisions, known as the Quark--Gluon Plasma (QGP). Measurements of the J/$\psi$ production serve as important tools for studying the properties of this medium. In addition to QGP effects, the J/$\psi$ production is modified by the presence of cold nuclear matter (CNM) effects such as shadowing or parton energy loss. These effects are studied in proton-nucleus collisions where no QGP formation is expected. In order to quantify how the J/$\psi$ production is modified by the medium, the vacuum production is modelled in proton-proton (pp) collisions and used as a reference for heavy-ion and proton-nucleus collisions. Measurements in pp collisions also serve as important tools for testing Quantum Chromodynamic (QCD) based models in both perturbative and non-perturbative regimes. 
The ALICE detector has the unique capability of measuring the J/$\psi$ at mid-rapidity through the di-electron channel down to zero transverse momentum. In this contribution a selection of the newest J/$\psi$ measurements at mid-rapidity performed by the ALICE Collaboration during the LHC Run 2 period will be discussed.  }

\FullConference{
European Physical Society Conference on High Energy Physics - EPS-HEP2019 -\\
			10-17 July, 2019\\
			Ghent, Belgium}

\begin{document}

\section{Physics motivations}
It is predicted that the extreme conditions of temperature and pressure created in (ultra)-relativistic heavy-ion collisions causes strongly interacting matter to undergo a phase transition into a plasma of deconfined quarks and gluons (QGP) \cite{Bjorken:1982qr}. During the initial hard parton-parton scattering heavy quarks, i.e. charm (c) and beauty (b), are produced and they therefore experience the full evolution of the QGP. Measurements of bound states of heavy quarks, such as the J/$\psi$, can therefore serve as sensitive probes of the strongly interacting medium. According to Ref. \cite{Matsui:1986dk}  the presence of free color charges in the deconfined medium causes a screening of the binding energy between quarks and anti-quarks suppressing the charmonium ($c\bar{c}$) production in heavy-ion collisions with respect to pp collisions. At high collision energies the production cross section of heavy quarks becomes large and it is therefore argued that a (re)generation of charmonium states takes place, either as a statistical recombination of uncorrelated heavy quark pairs at the phase boundary \cite{BraunMunzinger:2000px} or through the coalescence of charm quarks \cite{Thews:2000rj}.

The modification of charmonium production in nucleus-nucleus collisions includes contributions from Cold Nuclear Matter (CNM) effects, i.e. effects which arise in the absence of the QGP, such as shadowing, parton energy loss, and gluon saturation as e.g. in the Color Glass Condensate (CGC) \cite{NBrambilla2011}.  In order to quantify these effects, charmonium production in proton-lead collisions are studied. 

Measuring the charmonium production in proton-proton collisions constitutes the baseline for heavy-ion collisions. It also serves as a benchmark test for QCD-based production models. The production of charmonium states can be seen as a two-fold process with the initial creation of a charm pair evolving into a bound charmonium state. The initial creation can be described by perturbative QCD calculations, while the evolution into a bound state is a non-perturbative process. Currently none of the phenomenological models are able to successfully describe all the physical observables simultaneously. Thus, precise production measurements may contribute to better constraints on the production models and improve our understanding of how the charmonium states are produced.

\section{J/$\psi$ measurements at mid-rapidity in ALICE }
The ALICE detector \cite{Aamodt:2008zz} has a unique capability of measuring the J/$\psi$ at mid-rapidity down to zero transverse momentum (\pT). At mid-rapidity ($|y| < 0.9$) the J/$\psi$ is reconstructed through the di-electron decay channel with the central barrel. 
In the central barrel, the main tracking detectors are the Inner Tracking System (ITS) and the Time Projection Chamber (TPC). Situated around the beam line, the ITS provides information about the primary and secondary vertex position. The information of the secondary vertex position can be used to separate the contribution of J/$\psi$ mesons decaying from beauty hadrons (non-prompt). The TPC has full azimuthal coverage and provides particle identification (PID) based on the specific energy loss of charged particles traversing the detector with a resolution ranging from 5.5\% in pp events to 6.5\% in central Pb--Pb collisions. In particular the TPC provides good momentum resolution for electrons up to about 10~GeV/$c$.

\section{Results: selected highlights}
\begin{figure}
	\includegraphics[width=.5\textwidth]{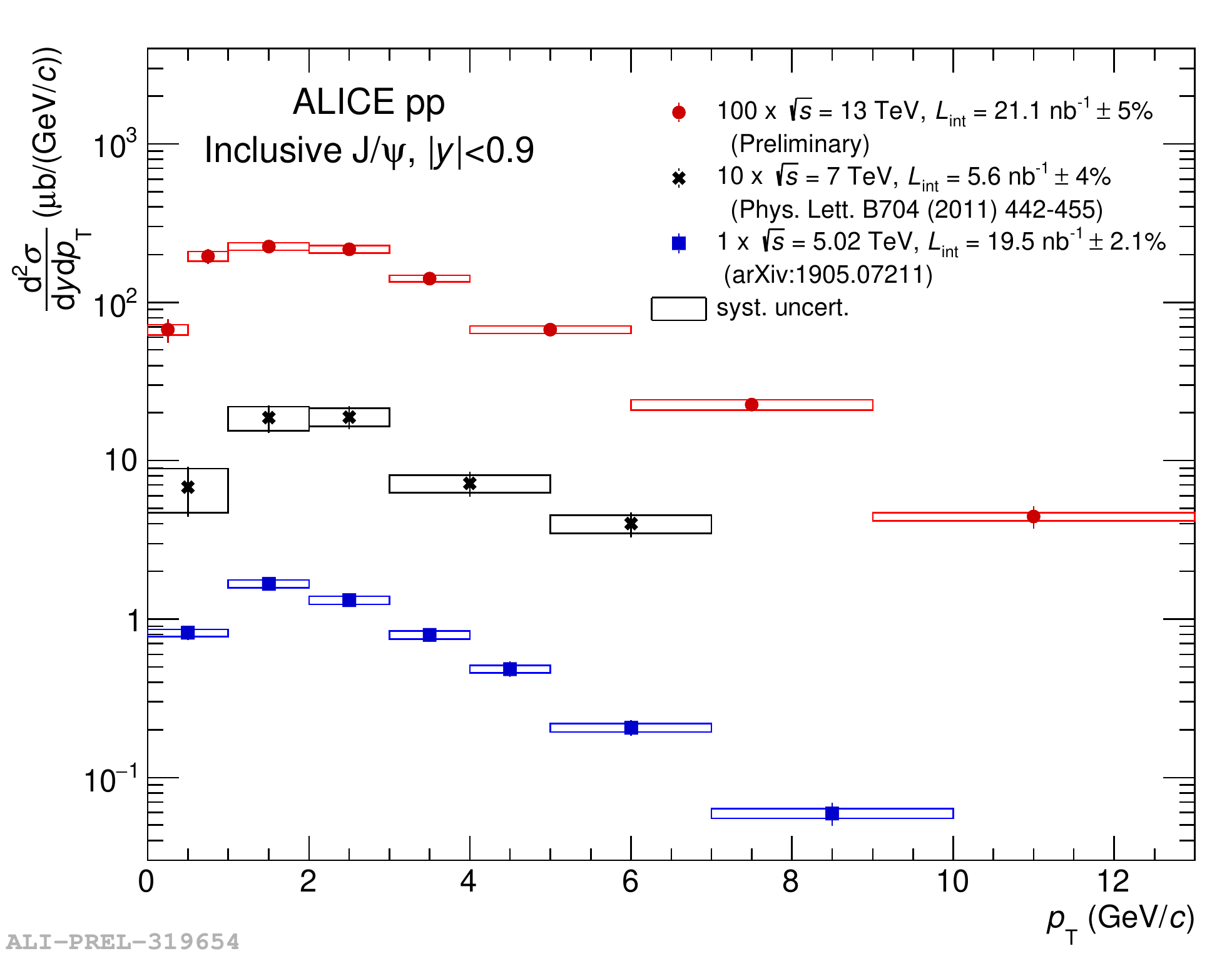} 
	\includegraphics[width=.5\textwidth]{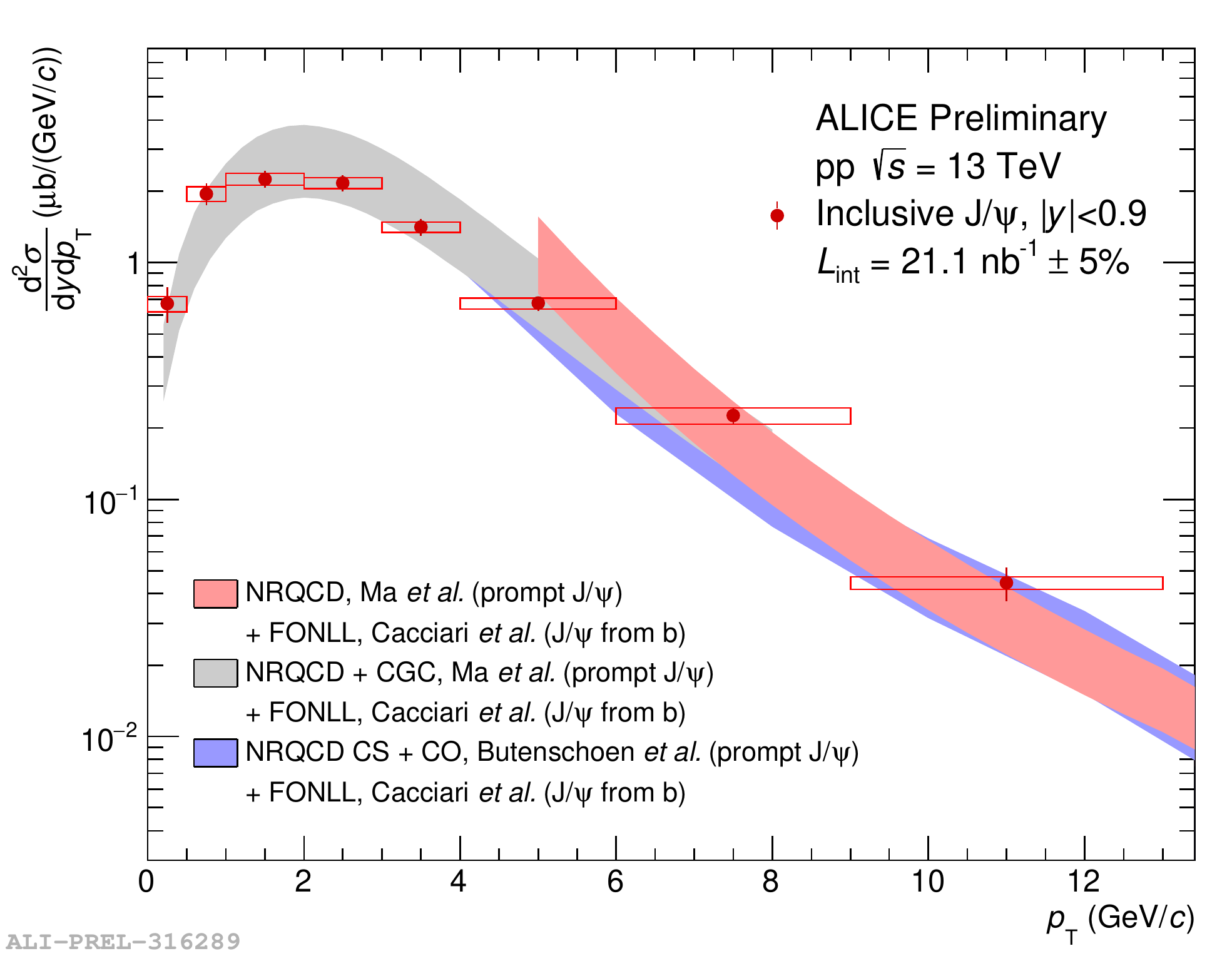}
	\caption{\textit{Left: }Inclusive J/$\psi$ cross section measurements at mid-rapidity in pp collisions for \sG \cite{Acharya:2019lkw} in red, \sA \cite{Aamodt:2011gj} in black, and preliminary results at \sD in blue, scaled by 1, 10, and 100, respectively.  \textit{Right:} \pT -differential inclusive J/$\psi$ cross section in pp collisions at \sD compared with pompt J/$\psi$ NRQCD  calculations \cite{Ma:2010yw,Butenschoen:2010rq,Ma:2014mri} combined with non-prompt J/$\psi$ calculations from FONLL \cite{Cacciari:2012ny}.
	}
	\label{fig:inclusiveCrossSection}
\end{figure}

\textbf{pp collisions:} The inclusive J/$\psi$ production cross section in pp collisions has been measured by the ALICE Collaboration at several center-of-mass energies. The left panel of Fig. \ref{fig:inclusiveCrossSection} shows the preliminary results for the inclusive J/$\psi$ cross section measured at \sD as a function of  transverse momentum (\pT) together with measurements at \sG \cite{Acharya:2019lkw} and \sA \cite{Aamodt:2011gj}. 

The right panel of Fig. \ref{fig:inclusiveCrossSection} shows the preliminary results for the inclusive J/$\psi$ cross section at \sD  as a function of \pT.
The measured cross section is compared to several Non-Relativistic QCD (NRQCD) calculations describing the contribution from prompt J/$\psi$ mesons \cite{Ma:2010yw,Butenschoen:2010rq,Ma:2014mri}, to which the non-prompt contribution calculated with Fixed-Order Next-To-Leading-Logarithm (FONLL) \cite{Cacciari:2012ny} has been added. The model by Ma \textit{et al.}  is coupled to a CGC description of the low-$x$ gluons in the proton providing a good description down to \pT = 0.  There is good agreement between the various model predictions and the measured cross section. It can be noted that at low \pT (up to about 4~GeV/$c$) the theoretical uncertainties are significantly larger than the uncertainties of the measured data points.

\begin{figure}
\center
	\includegraphics[width=.45\textwidth]{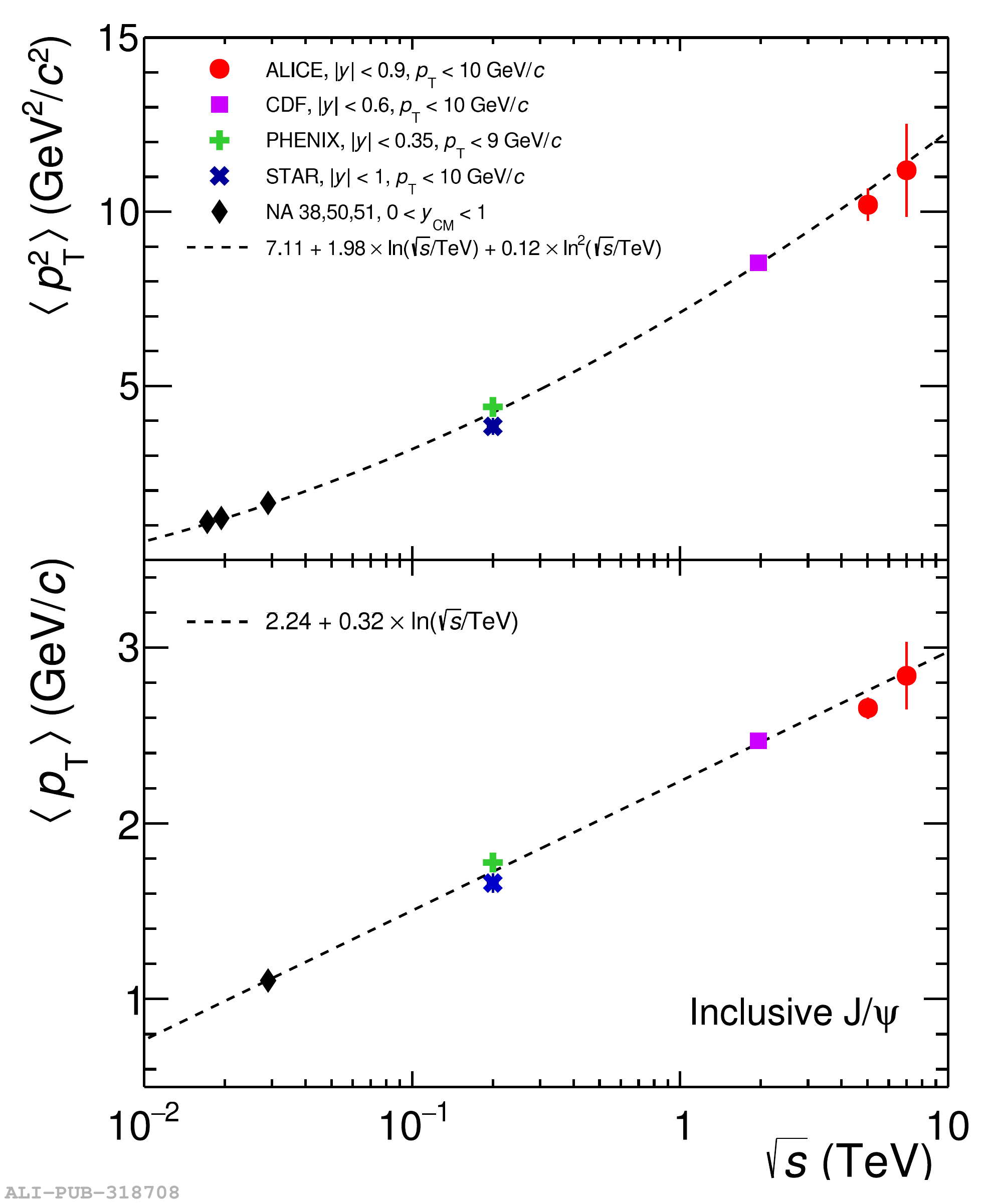}

	\caption{ The inclusive J/$\psi$ $\langle p_{\mbox{\tiny T}} \rangle$ (upper panel) and $\langle p_{\mbox{\tiny T}}^{2} \rangle$ (lower panel) in pp collisions as a function of collision energy  (see Ref. \cite{Acharya:2019lkw} and references therein).   }
	\label{fig:energydep}
\end{figure}

The center-of-mass energy (\s) dependence of the \pT -differential cross section has been studied by looking at the average transverse momentum $\langle p_{\mbox{\tiny T}} \rangle$ and the squared average transverse momentum $\langle p_{\mbox{\tiny T}}^{2} \rangle$ (see Ref. \cite{Acharya:2019lkw} and references therein). The inclusive J/$\psi$ $\langle p_{\mbox{\tiny T}} \rangle$ and $\langle p_{\mbox{\tiny T}}^{2} \rangle$ are determined by a fit to the \pT spectrum using a power law function of the form

\begin{equation}
f \left( p_{\mbox{\tiny T}} \right) = C \times \frac{ p_{\mbox{\tiny T}}}{\left\{ 1 + \left( p_{\mbox{\tiny T}}/p_0\right)^2\right\}^n}  \, ,
\end{equation}
 where $C$, $p_0$, and $n$ are free fit parameters. 
Figure \ref{fig:energydep} shows the inclusive J/$\psi$ $\langle p_{\mbox{\tiny T}} \rangle$ and $\langle p_{\mbox{\tiny T}}^{2} \rangle$ as a function of \s where a steady increase with energy can be observed. The measurements are described well using a linear and squared logarithmic increase with energy, respectively. The rising energy dependence originates from the fact that for a given Bjorken $x$, the momentum transfer between the scattering partons increases with energy, leading to a hardening of the resulting \pT spectrum. 

\begin{figure}
	\includegraphics[width=.53\textwidth]{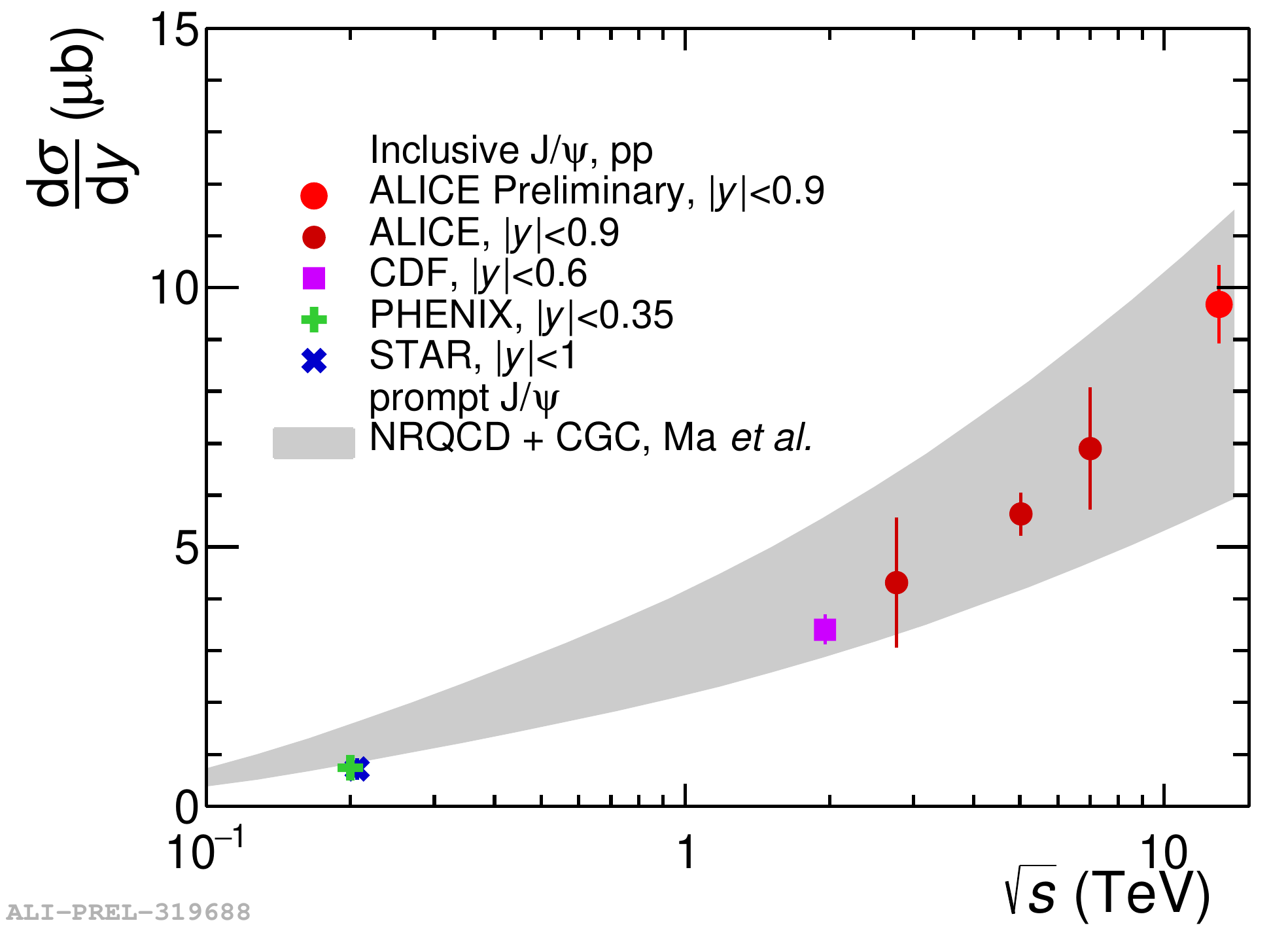}
	\includegraphics[width=.47\textwidth]{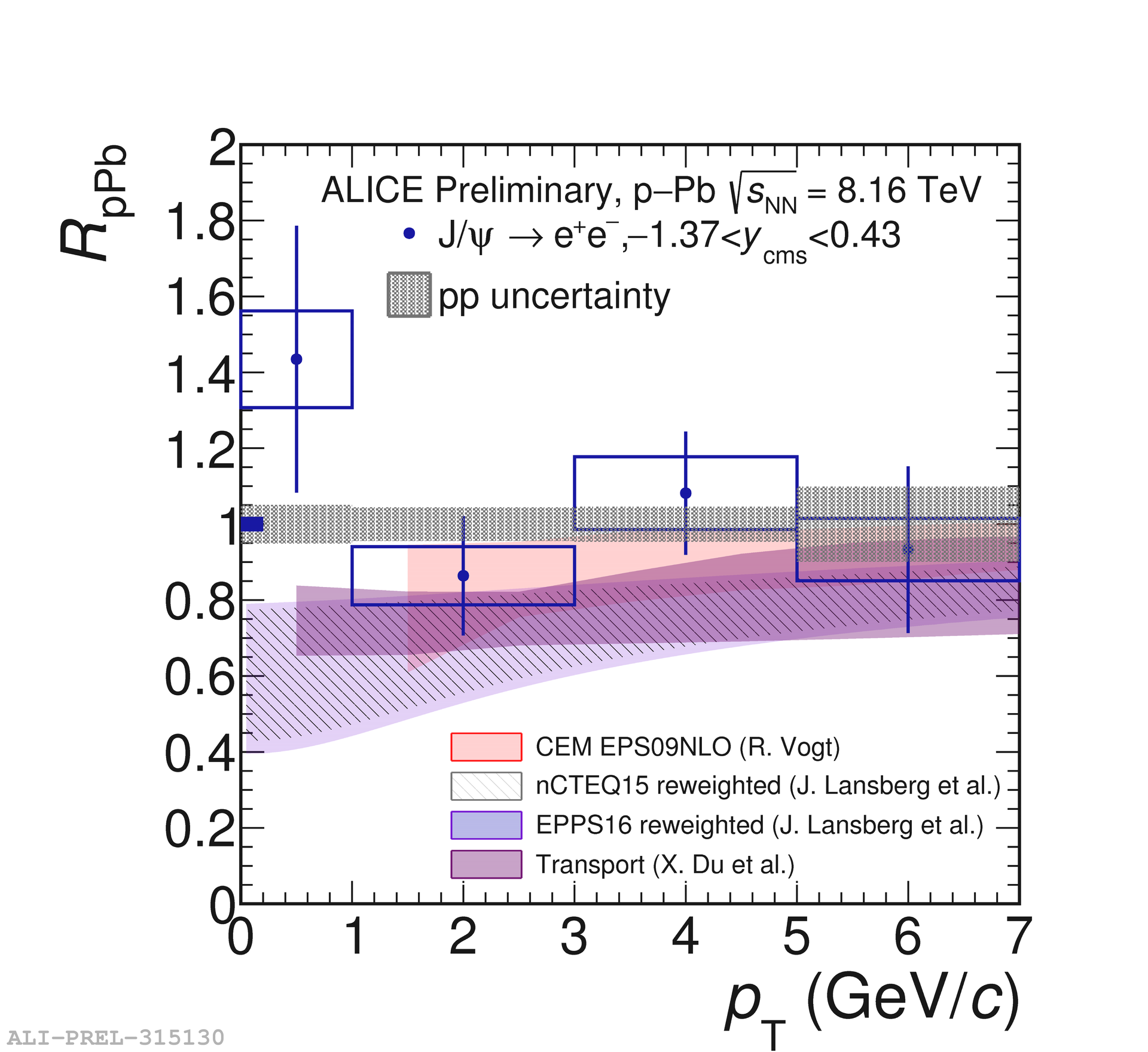}
	\caption{\textit{Right:} Energy dependence of the inclusive J/$\psi$ cross section in pp collisions compared with CGC + NRQCD model calculations from \cite{Ma:2014mri}. The preliminary results at \sD are compared with measurements from PHENIX, STAR, CDF and ALICE (See \cite{Acharya:2019lkw} and references therein). The data points from PHENIX and STAR are slightly shifted for visibility. \textit{Left:} $R_\text{pPb}$  of inclusive J/$\psi$  as a function of \pT in p--Pb collsions at \sU. The results are compared to several model calculations \cite{Albacete:2017qng,Lansberg:2016deg,Kusina:2017gkz,Du:2018wsj}.}
	\label{fig:pPb}
\end{figure}

The left panel of Fig. \ref{fig:pPb} shows the \pT -integrated inclusive J/$\psi$ cross section as a function of \s (see Ref. \cite{Acharya:2019lkw} and references therein). The error bars represent the quadratic sum of the statistical and systematic uncertainties including the uncertainty of the luminosity. The measurements are compared to a low-$x$ CGC + NRQCD \cite{Ma:2014mri} prediction for prompt J/$\psi$. The contribution from non-prompt J/$\psi$ is less than 10\% for the \pT -integrated cross section and increases with increasing energy. Taking this into account the predicted energy dependence agrees well with the data, but with significant uncertainties.

\textbf{p--Pb:} Modifications of the J/$\psi$ production in p--Pb collisions are quantified by the nuclear modification factor
\begin{equation}
	R_\text{pPb} = \frac{\sigma_\text{pPb}}{A_\text{pPb}\sigma_\text{pp}} \, ,
\end{equation}
where the measured charmonium cross section in p--Pb collisions $\sigma_\text{pPb}$ is normalized by the measured cross section in pp collisions $\sigma_\text{pp}$ at the same collision energy scaled by the atomic number of the lead nucleus $A_\text{pPb}$. In the absence of nuclear effects, $R_\text{pPb}$ is expected to be unity.  

The right panel of Fig. \ref{fig:pPb} shows the preliminary $R_\text{pPb}$  of inclusive J/$\psi$ at \sU as a function of \pT. The measurement is compared to several calculations that include only shadowing as a cold nuclear matter effect \cite{Albacete:2017qng,Lansberg:2016deg,Kusina:2017gkz} and a model including also contributions from the final state interactions between $c\bar{c}$ pairs and the partonic/hadronic systems created in the collisions \cite{Du:2018wsj}. In the latter model the nuclear shadowing still plays the dominant role in determining the values of the nuclear modification factors.
The measured data and model predictions are compatible within the current uncertainties.

\textbf{Pb--Pb:} The nuclear modification factor in Pb--Pb collisions for a given centrality class $i$ is calculated as
\begin{equation}
	R^{i}_\text{PbPb} \left(p_{\mbox{\tiny T}} \right) = \frac{ \text{d} N^{i}_\text{PbPb}/\text{d} p_{\mbox{\tiny T}}}{\langle T^{i}_\text{AA}\rangle \times \text{d}\sigma^{i}_\text{pp}/\text{d}p_{\mbox{\tiny T}}} \, ,
\end{equation}
where $\text{d}N^{i}_\text{PbPb}/\text{d}p_{\mbox{\tiny T}}$ is the corrected yield in the Pb--Pb collisions normalized to the measured yield in pp collisions at the same center-of-mass energy scaled by the nuclear overlap function, $\langle T^{i}_\text{AA}\rangle$. 

\begin{figure}
	\includegraphics[width=.5\textwidth]{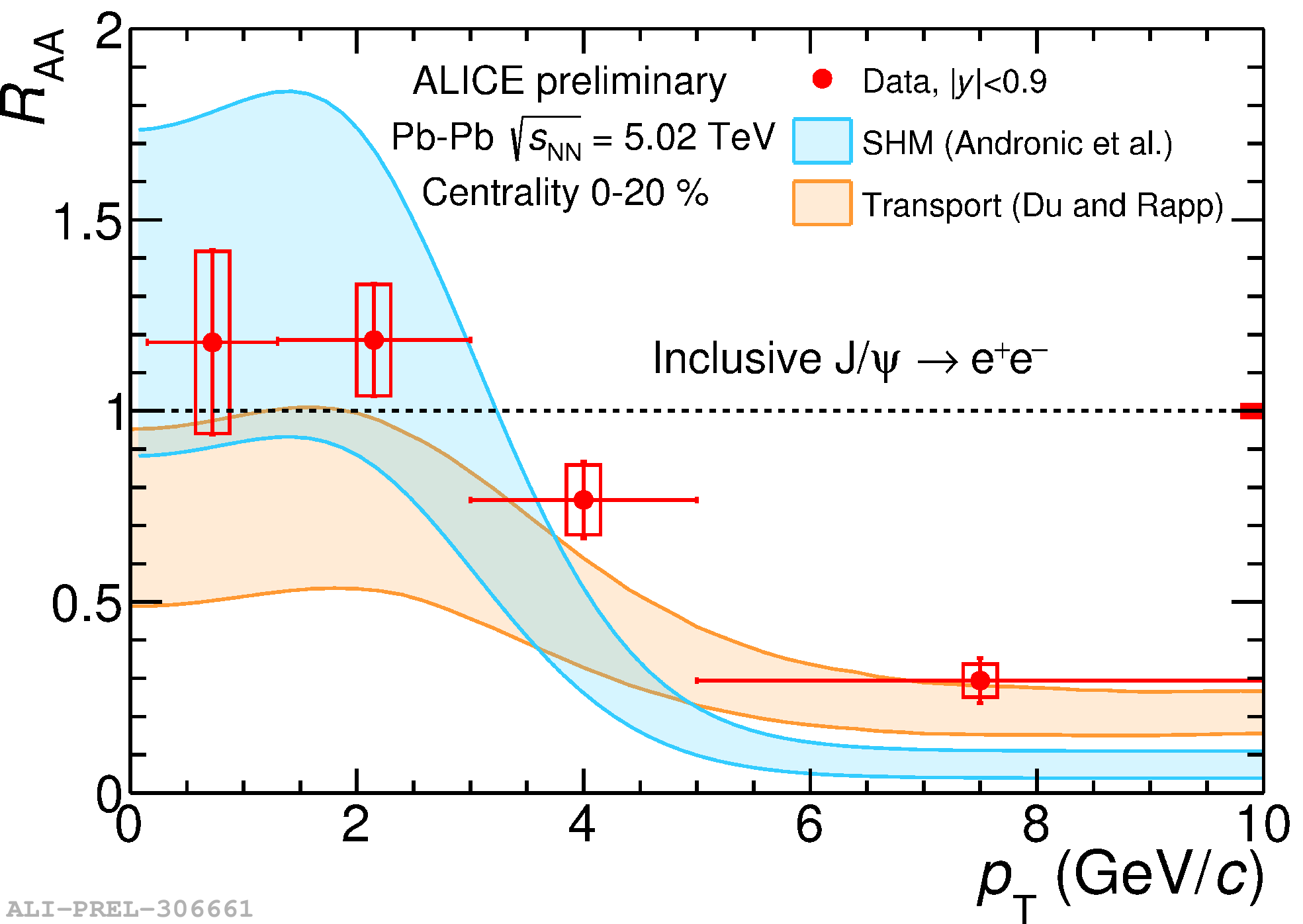}
	\includegraphics[width=.5\textwidth]{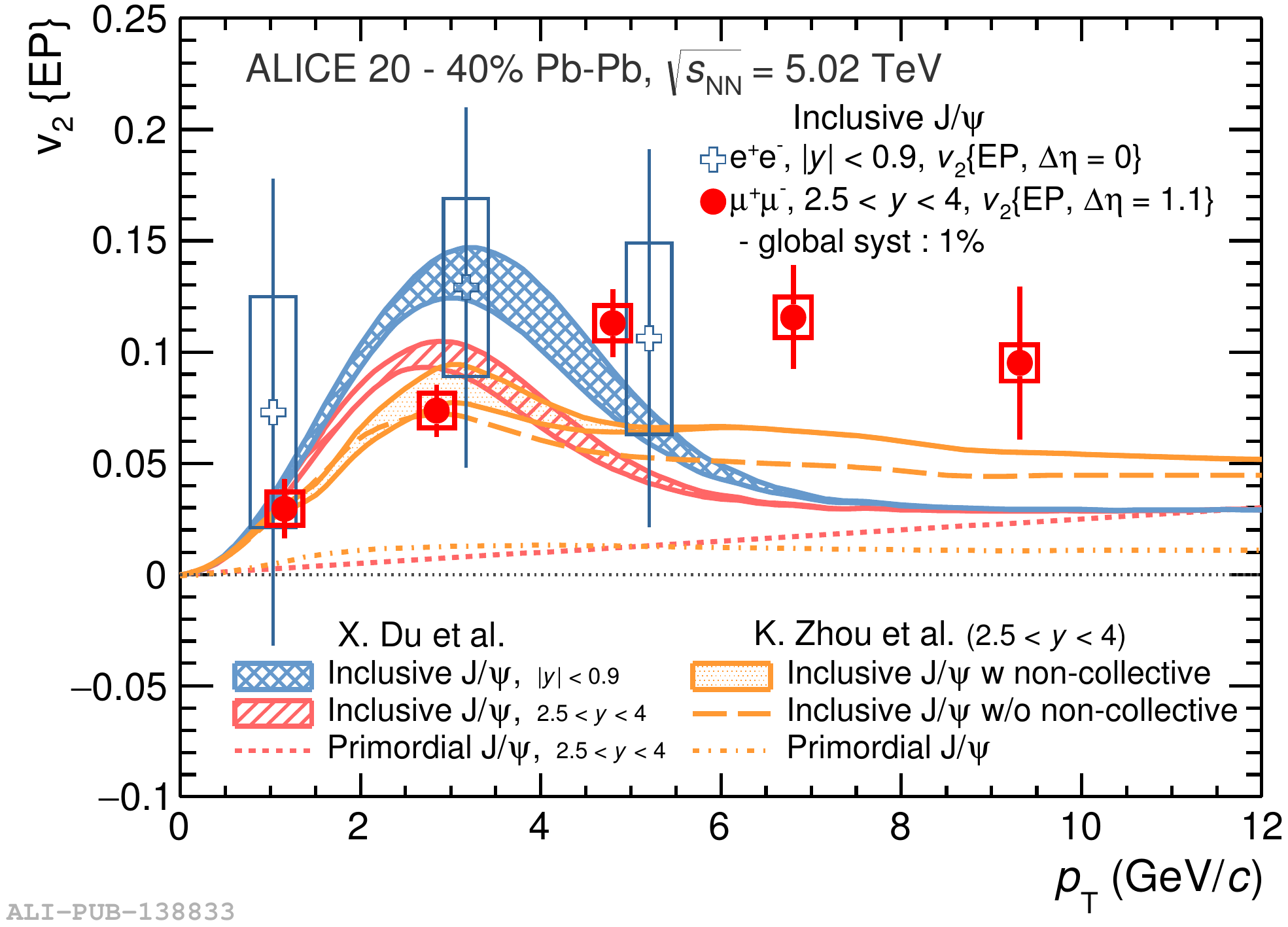}
	\caption{ \textit{Left:} Inclusive J/$\psi$ $R_\text{AA}$ with centrality 0-20\% as a function of \pT in Pb--Pb collisions at \sC. The results are compared to statistical hadronization and transport models \cite{Andronic:2016nof,Du:2015wha}.  \textit{Right:} Inclusive J$\psi$ $v_2$ (\pT) at forward and mid-rapidity for semicentral (20\% - 40\%) Pb--Pb collisions at \sG with transport model calculations (See \cite{Acharya:2017tgv} and references therein.). }
	\label{fig:PbPb}
\end{figure}

The left panel of Fig. \ref{fig:PbPb} shows the preliminary $R_\text{AA}$ of the inclusive J/$\psi$  at \sC with centrality 0-20\% as a function of \pT in Pb--Pb collisions. An increase in the suppression is observed with increasing \pT. The measured $R_\text{AA}$ is compared to two different models. The statistical hadronization model \cite{Andronic:2016nof} assumes that the J/$\psi$ formation only happens at the chemical freeze-out based on their statistical weights, while the transport model \cite{Du:2015wha} predicts continuous dissociation and regeneration both inside the QGP and the hadronic phase. Both models describe the measured trend within the current experimental and theoretical uncertainties. The large model uncertainties originate from the choice of input parameters, in particular the charm production cross section and the set of nuclear parton distribution functions (nPDFs).

The second harmonic of the J/$\psi$ momentum azimuthal distribution with respect to the collisions reaction plane ($v_2$) quantifies the elliptic flow  and is sensitive to the geometry and the dynamics of the early stages of the collisions. The J/$\psi$ $v_2$ in Pb--Pb collisions at \sC is shown in the right panel of Fig. \ref{fig:PbPb} measured at both forward- and mid-rapidity (see Ref. \cite{Acharya:2017tgv} and references therein). In both rapidity regions there is evidence of a positive flow. Transport models implementing (re)generation components describe both measurements at low \pT indicating that (re)combined J/$\psi$ inherit the elliptic flow from thermalized charm quarks. However, at high \pT the models underestimate the measured flow. The origin for this mismatch is currently not understood suggesting that the models are missing some mechanism. 

\section{Conclusions and perspectives}
Selected studies of J/$\psi$ production at mid-rapidity in pp, p--Pb, and Pb--Pb collisions performed by the ALICE collaboration are presented. In pp collisions, precise differential cross section measurements extended down to \pT = 0 are shown at several center-of-mass energies. NRQCD + CGC calculations provide a fair description of the measured data. A hardening of the \pT -differential cross section is observed with increasing collision energy.
In p--Pb the measured $R_\text{pPb}$ is compatible with model predictions within the current uncertainties.
The measured $R_\text{AA}$ in Pb--Pb collisions shows that the suppression increases with increasing \pT. 
This is consistent with a significant contribution to the J/$\psi$ yield from the (re)generation mechanism. However, a higher precision is needed both for measurements and the theoretical predictions in order to distinguish between various models.
Indications of a positive J/$\psi$ $v_2$ both at forward and mid-rapidity in Pb--Pb collisions suggests thermalization of charm quarks within the medium. At high \pT the different theoretical predictions underestimate the measured flow indicating a missing mechanism in the models.

A significant improvement in the statistical uncertainties of the J/$\psi$ measurements is expected for the LHC Run 3 (starting in 2021) and Run 4 due to the scheduled upgrade of the ALICE detector \cite{Citron:2018lsq}. Moving to continuous readout will result in a high statistics minimum bias sample which for Pb--Pb collisions is expected to give $\mathcal{L}_\text{int} = 10~\text{nb}^{-1}$. Upgrades on the ITS will improve the tracking and vertex precision
 increasing significantly the ability to separate prompt and non-prompt J/$\psi$. This will lead to precise measurements of their respective nuclear modification factors and elliptic flow, down to almost zero \pT. Larger experimental data samples and improved detector performance will allow for new measurements at mid-rapidity, such as the $\psi$(2S) yielding very low signal to background ratios.

\bibliographystyle{JHEP}
\bibliography{BibliographyNew.bib}

\providecommand{\href}[2]{#2}\begingroup\raggedright\begin{thebibliography}{10}

\bibitem{Bjorken:1982qr}
J.~D. Bjorken, \emph{{Highly Relativistic Nucleus-Nucleus Collisions: The
  Central Rapidity Region}},
  \href{https://doi.org/10.1103/PhysRevD.27.140}{\emph{Phys. Rev.} {\bfseries
  D27} (1983) 140}.

\bibitem{Matsui:1986dk}
T.~Matsui and H.~Satz, \emph{{$J/\psi$ Suppression by Quark-Gluon Plasma
  Formation}}, \href{https://doi.org/10.1016/0370-2693(86)91404-8}{\emph{Phys.
  Lett.} {\bfseries B178} (1986) 416}.

\bibitem{BraunMunzinger:2000px}
P.~Braun-Munzinger and J.~Stachel, \emph{{(Non)thermal aspects of charmonium
  production and a new look at J / psi suppression}},
  \href{https://doi.org/10.1016/S0370-2693(00)00991-6}{\emph{Phys. Lett.}
  {\bfseries B490} (2000) 196}
  [\href{https://arxiv.org/abs/nucl-th/0007059}{{\ttfamily nucl-th/0007059}}].

\bibitem{Thews:2000rj}
R.~L. Thews, M.~Schroedter and J.~Rafelski, \emph{{Enhanced $J/\psi$ production
  in deconfined quark matter}},
  \href{https://doi.org/10.1103/PhysRevC.63.054905}{\emph{Phys. Rev.}
  {\bfseries C63} (2001) 054905}
  [\href{https://arxiv.org/abs/hep-ph/0007323}{{\ttfamily hep-ph/0007323}}].

\bibitem{NBrambilla2011}
N.~Brambilla. et~al., \emph{{Heavy quarkonium: progress,puzzles, and
  opportunities}},
  \href{https://doi.org/10.1140/epjc/s10052-010-1534-9}{\emph{Eur.Phys.J.}
  {\bfseries C71} (2011) 1534}
  [\href{https://arxiv.org/abs/1010.5827}{{\ttfamily 1010.5827}}].

\bibitem{Aamodt:2008zz}
{\scshape ALICE} collaboration, \emph{{The ALICE experiment at the CERN LHC}},
  \href{https://doi.org/10.1088/1748-0221/3/08/S08002}{\emph{JINST} {\bfseries
  3} (2008) S08002}.

\bibitem{Acharya:2019lkw}
{\scshape ALICE} collaboration, \emph{{Inclusive J/$\psi$ production at
  mid-rapidity in pp collisions at $\sqrt{s}$ = 5.02 TeV}},
  \href{https://doi.org/10.1007/JHEP10(2019)084}{\emph{JHEP} {\bfseries 10}
  (2019) 084} [\href{https://arxiv.org/abs/1905.07211}{{\ttfamily
  1905.07211}}].

\bibitem{Aamodt:2011gj}
{\scshape ALICE} collaboration, \emph{{Rapidity and transverse momentum
  dependence of inclusive J$/\psi$ production in $pp$ collisions at $\sqrt{s} =
  7$ TeV}}, \href{https://doi.org/10.1016/j.physletb.2011.09.054,
  10.1016/j.physletb.2012.10.060}{\emph{Phys. Lett.} {\bfseries B704} (2011)
  442} [\href{https://arxiv.org/abs/1105.0380}{{\ttfamily 1105.0380}}].

\bibitem{Ma:2010yw}
Y.-Q. Ma, K.~Wang and K.-T. Chao, \emph{{$J/\psi (\psi^\prime)$ production at
  the Tevatron and LHC at ${\cal O}(\alpha_s^4v^4)$ in nonrelativistic QCD}},
  \href{https://doi.org/10.1103/PhysRevLett.106.042002}{\emph{Phys. Rev. Lett.}
  {\bfseries 106} (2011) 042002}
  [\href{https://arxiv.org/abs/1009.3655}{{\ttfamily 1009.3655}}].

\bibitem{Butenschoen:2010rq}
M.~Butenschoen and B.~A. Kniehl, \emph{{Reconciling $J/\psi$ production at
  HERA, RHIC, Tevatron, and LHC with NRQCD factorization at next-to-leading
  order}}, \href{https://doi.org/10.1103/PhysRevLett.106.022003}{\emph{Phys.
  Rev. Lett.} {\bfseries 106} (2011) 022003}
  [\href{https://arxiv.org/abs/1009.5662}{{\ttfamily 1009.5662}}].

\bibitem{Ma:2014mri}
Y.-Q. Ma and R.~Venugopalan, \emph{{Comprehensive Description of J/$\psi$
  Production in Proton-Proton Collisions at Collider Energies}},
  \href{https://doi.org/10.1103/PhysRevLett.113.192301}{\emph{Phys. Rev. Lett.}
  {\bfseries 113} (2014) 192301}
  [\href{https://arxiv.org/abs/1408.4075}{{\ttfamily 1408.4075}}].

\bibitem{Cacciari:2012ny}
M.~Cacciari, S.~Frixione, N.~Houdeau, M.~L. Mangano, P.~Nason and G.~Ridolfi,
  \emph{{Theoretical predictions for charm and bottom production at the LHC}},
  \href{https://doi.org/10.1007/JHEP10(2012)137}{\emph{JHEP} {\bfseries 10}
  (2012) 137} [\href{https://arxiv.org/abs/1205.6344}{{\ttfamily 1205.6344}}].

\bibitem{Albacete:2017qng}
J.~L. Albacete et~al., \emph{{Predictions for Cold Nuclear Matter Effects in
  $p+$Pb Collisions at $\sqrt{s_{_{NN}}} = 8.16$ TeV}},
  \href{https://doi.org/10.1016/j.nuclphysa.2017.11.015}{\emph{Nucl. Phys.}
  {\bfseries A972} (2018) 18}
  [\href{https://arxiv.org/abs/1707.09973}{{\ttfamily 1707.09973}}].

\bibitem{Lansberg:2016deg}
J.-P. Lansberg and H.-S. Shao, \emph{{Towards an automated tool to evaluate the
  impact of the nuclear modification of the gluon density on quarkonium, D and
  B meson production in proton-nucleus collisions}},
  \href{https://doi.org/10.1140/epjc/s10052-016-4575-x}{\emph{Eur. Phys. J.}
  {\bfseries C77} (2017) 1} [\href{https://arxiv.org/abs/1610.05382}{{\ttfamily
  1610.05382}}].

\bibitem{Kusina:2017gkz}
A.~Kusina, J.-P. Lansberg, I.~Schienbein and H.-S. Shao, \emph{{Gluon Shadowing
  in Heavy-Flavor Production at the LHC}},
  \href{https://doi.org/10.1103/PhysRevLett.121.052004}{\emph{Phys. Rev. Lett.}
  {\bfseries 121} (2018) 052004}
  [\href{https://arxiv.org/abs/1712.07024}{{\ttfamily 1712.07024}}].

\bibitem{Du:2018wsj}
X.~Du and R.~Rapp, \emph{{In-Medium Charmonium Production in Proton-Nucleus
  Collisions}}, \href{https://doi.org/10.1007/JHEP03(2019)015}{\emph{JHEP}
  {\bfseries 03} (2019) 015}
  [\href{https://arxiv.org/abs/1808.10014}{{\ttfamily 1808.10014}}].

\bibitem{Andronic:2016nof}
A.~Andronic, P.~Braun-Munzinger, K.~Redlich and J.~Stachel, \emph{{Hadron
  yields, the chemical freeze-out and the QCD phase diagram}},
  \href{https://doi.org/10.1088/1742-6596/779/1/012012}{\emph{J. Phys. Conf.
  Ser.} {\bfseries 779} (2017) 012012}
  [\href{https://arxiv.org/abs/1611.01347}{{\ttfamily 1611.01347}}].

\bibitem{Du:2015wha}
X.~Du and R.~Rapp, \emph{{Sequential Regeneration of Charmonia in Heavy-Ion
  Collisions}},
  \href{https://doi.org/10.1016/j.nuclphysa.2015.09.006}{\emph{Nucl. Phys.}
  {\bfseries A943} (2015) 147}
  [\href{https://arxiv.org/abs/1504.00670}{{\ttfamily 1504.00670}}].

\bibitem{Acharya:2017tgv}
{\scshape ALICE} collaboration, \emph{{J/$\psi$ elliptic flow in Pb-Pb
  collisions at $\sqrt{s_\mathrm{NN}}=5.02$ TeV}},
  \href{https://doi.org/10.1103/PhysRevLett.119.242301}{\emph{Phys. Rev. Lett.}
  {\bfseries 119} (2017) 242301}
  [\href{https://arxiv.org/abs/1709.05260}{{\ttfamily 1709.05260}}].

\bibitem{Citron:2018lsq}
Z.~Citron et~al., \emph{{Future physics opportunities for high-density QCD at
  the LHC with heavy-ion and proton beams}},  in \emph{{HL/HE-LHC Workshop:
  Workshop on the Physics of HL-LHC, and Perspectives at HE-LHC Geneva,
  Switzerland, June 18-20, 2018}}, 2018,
  \href{https://arxiv.org/abs/1812.06772}{{\ttfamily 1812.06772}}.

\end{thebibliography}\endgroup


\end{document}